\documentclass[prb,twocolumn,showpacs,floatfix]{revtex4}
\usepackage{graphics,epsfig,amsfonts,amssymb,amsmath,color}
\begin{document}
\title{Orbital differentiation and the role of orbital ordering in the magnetic state of Fe superconductors}
\author{E. Bascones}
\email{leni.bascones@icmm.csic.es}
\affiliation{Instituto de Ciencia de Materiales de Madrid,
ICMM-CSIC, Cantoblanco, E-28049 Madrid (Spain).}
\author{B. Valenzuela}
\email{belenv@icmm.csic.es}
\affiliation{Instituto de Ciencia de Materiales de Madrid,
ICMM-CSIC, Cantoblanco, E-28049 Madrid (Spain).}
\author{M.J. Calder\'on}
\email{calderon@icmm.csic.es}
\affiliation{Instituto de Ciencia de Materiales de Madrid,
ICMM-CSIC, Cantoblanco, E-28049 Madrid (Spain).}

\date{\today}
\begin{abstract}
We analyze the metallic $(\pi,0)$ antiferromagnetic state of a 
five-orbital model for iron superconductors. We find that with increasing 
interactions the system does not evolve trivially from the pure itinerant to 
the pure localized regime. Instead we find a region with a strong orbital differentiation between $xy$ 
and $yz$, which are half-filled gapped states at the Fermi level, and  itinerant $zx$, $3z^2-r^2$ and $x^2-y^2$. 
We argue that orbital ordering between the $yz$ and $zx$ orbitals arises 
as a consequence of the interplay of the exchange energy in the 
antiferromagnetic $x$ direction and 
the kinetic energy gained by the itinerant orbitals along the ferromagnetic 
$y$ direction with an overall dominance of the kinetic energy gain. 
We indicate that iron superconductors may be close to the boundary
between the itinerant and the orbital differentiated regimes and that it could be possible to cross this boundary with doping.    
 
\end{abstract}
\pacs{75.10.Jm, 75.10.Lp, 75.30.Ds}
\maketitle


There is a strong interrelation between the orbital degree of freedom, 
the magnetization, and the lattice structure in the Fe-superconductors. 
Unveiling the nature of these connections would define the landscape from which 
superconductivity emerges in these materials. One of the important issues is 
the determination of the strength of the interactions. Unlike the cuprates, 
which are antiferromagnetic Mott insulators when undoped, the 
Fe-superconductors are antiferromagnetic metals, highlighting the relevance of 
the itinerancy of the conduction electrons. 
Undoped materials have to accomodate six electrons in the
five Fe-d orbitals, with an  average filling of $1.2$,
close to the one of doped Mott insulators.\cite{liebsch2010,Werner2012,imadaPRL2012}
The itinerant\cite{raghu08,mazin08-2,chubukov08,cvetkovic09} versus localized\cite{yildirim08,si08}
origin of the magnetization has been discussed since the discovery of 
superconductivity in these systems.

On the other hand there is increasing evidence for orbital differentiation 
and a possible coexistence of itinerant and localized 
electrons.\cite{Castro-Neto,demedici2009,lv_phillips10,yin10} 
Angle Resolved Photoemission Spectroscopy (ARPES) measurements report different 
renormalization values for the various bands close to the Fermi energy 
depending on their orbital character.\cite{shen2012,sudayama2012} 
Similar qualitative conclusions may be 
inferred from Dynamical Mean Field Theory (DMFT) and slave-spin 
calculations.\cite{liebsch2010,aichhorn2010,yin11,si2012}  

The possible role of orbital ordering in the magnetism 
is of present interest. 
The current debate is focused on whether it is the leading instability driving the 
magnetism\cite{phillips09} or it appears as a consequence of the 
magnetic ordering,\cite{leeyinku09,kruger09,nosotrasprl10,Dagotto10arpes} 
as well as its possible relation to the observed
anisotropic properties.\cite{chu10-1,mazin10,nosotrasprl10_2,chen_deveraux10,degiorgi10,zhaonatphys09,singh-09,shimojima10,sciencedavis10,uchida2011,lv-phillips2011,fernandes2011,yin11,fisher2011}
In particular, the resistivity in the $(\pi,0)$ antiferromagnetic state was 
measured to be larger in the ferromagnetic $y$-direction than in the 
antiferromagnetic $x$-direction\cite{chu10-1,mazin10} with a change in sign 
upon hole doping.\cite{sign-reversal2012}

\begin{figure}
\leavevmode
\includegraphics[clip,width=0.47\textwidth]{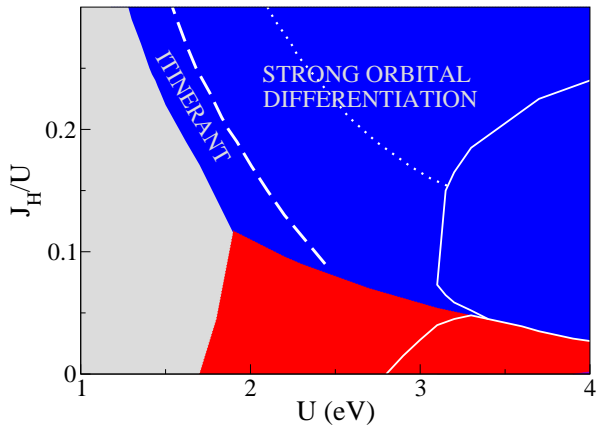}
\caption{
(Color online) This figure summarizes the main results of this work. The colors distinguish the different $(\pi,0)$ magnetic phases of the undoped ($6$ electrons in $5$ d orbitals) system as a function of the local interaction parameters 
$U$ and $J_H$.\cite{nosotrasprl10}   The grey area is the non-magnetic region. 
The blue and red areas are magnetic with a high moment (parallel orbital 
moments) and a low moment (antiparallel orbital moments) state respectively. 
The white solid lines on the right separate the metallic ($U\lesssim 3$) from 
the insulating ($U \gtrsim 3$) state.  We have analyzed the orbital 
differentiation within the blue metallic area. We distinguish two different regions that we label, with increasing $U$, as {\it itinerant} and {\it strong orbital differentiation}. The strong orbital differentiation region can be further splitted by the opening of a gap at half-filling, see text for discussion.
}
\label{fig:pd} 
\end{figure}

In order to shed light on the role of the different orbitals on the magnetic 
state of Fe-superconductors, we analyze the metallic $(\pi,0)$ 
antiferromagnetic state as a function of the interactions treated within mean-field. 
 Close to the 
non-magnetic phase boundary, electrons are itinerant. An $S=2$ state compatible 
with a localized $J_1-J_2$ description is found deep in the insulating 
regime.\cite{nosotrasprl10} 
With increasing interactions the system does not 
evolve trivially from the pure itinerant to the pure localized regime. 
Instead we find a region with a strong orbital differentiation,
see Fig.~\ref{fig:pd}. 
In this region, $zx$, $3z^2-r^2$ and $x^2-y^2$ are itinerant while $xy$ 
and $yz$ are half-filled and have a gap at the Fermi level. These gapped 
states are reminiscent of the localized electrons discussed in the literature.\cite{Castro-Neto,demedici2009,yin10,shen-orb-selective2012} 
At large values of Hund's coupling the itinerant 
electrons are also gapped at half-filling while keeping a finite density at the Fermi level. 
We uncover the different role that orbitals play in the stabilization of 
the orbital ordering and consequently of the $(\pi,0)$ antiferromagnetic state. 
Within a model of localized and itinerant orbitals, we find that while superexchange\cite{leeyinku09} between $xy$ and $yz$ contributes to generate orbital ordering between $yz$ 
and $zx$,  it is necessary to invoke the kinetic energy gain of the 
itinerant electrons along the ferromagnetic direction to describe the 
observed features.
We analyze this result in connection with the resistivity anisotropy.\cite{nosotrasprl10_2,chen_deveraux10}
We argue that iron pnictides are close to the boundary between {\it itinerant} 
and 
{\it strong orbital differentiation} regimes and that it could be  
possible to cross this 
boundary with doping.


We consider an interacting two-dimensional five-orbital model for the FeAs layer, as 
described in Ref.~[\onlinecite{nosotrasprl10}]. The Fe orbitals are defined 
within the one-iron unit cell and hence $x$ and $y$ are given by the Fe-Fe 
nearest neighbor directions. Only local interactions are included. 
Interactions with rotational symmetry 
can be expressed
in terms of only two parameters: the intra-orbital Hubbard $U$ and the 
Hund's coupling $J_H$.\cite{castellani78}  
We focus on the metallic $(\pi,0)$ antiferromagnetic state and study 
the phase diagram as a function of 
$U$ and $J_H/U$ with interactions treated at the Hartree-Fock level.\cite{nosotrasprl10}

For the tight-binding we use the model described in Ref.~[\onlinecite{nosotrasprb09}]
and obtained within the Slater-Koster 
formalism\cite{slater54} that takes into account the symmetry of the orbitals and the lattice.
In this model the tight-binding parameters
are analytic functions of the angle $\alpha$ formed by the Fe-As bond and 
the Fe plane. The resulting bands, their orbital compositions, 
the Fermi surface, and the modifications induced by $\alpha$ are 
all consistent with ab-initio calculations.\cite{nosotrasprb09} 
This allows us to straightforwardly explore the effect of the Fe-plane 
lattice structure on the electronic properties.
Except when specifically stated,
the results are obtained for the undoped case with six electrons per iron
$n=6$, and regular tetrahedra $\alpha=35.3^o$.

The main results of our analysis are summarized in Fig.~\ref{fig:pd} on top of 
the $(\pi,0)$ magnetic phase diagram previously reported in 
Ref.~[\onlinecite{nosotrasprl10}]. 
The system becomes insulating on the right of 
the solid white lines. 
In the grey region the system is not magnetic and the red area corresponds to a low moment
state, in which Hund's rule is violated.\cite{nosotrasprl10,cricchio09,liu11,schikling2011}
The blue high moment state fulfills Hund's rule.
Deep in the insulating regime, this 
state has a spin $S=2$ with filled $x^2-y^2$ and half-filled $xy$, $yz$, $zx$, and $3z^2-r^2$~[\onlinecite{nosotrasprl10}].
In the region of larger $U$, an increasing $J_H/U$ leads to 
metallicity.\cite{nosotrasprb2012} 
We focus here on the metallic blue 
state, on the left of the solid line. 
Decreasing the value of $U$ we find 
different regions which differ on 
their electronic structure and the related orbital differentiation. 
The regions have been labelled {\it strong orbital differentiation} and 
{\it itinerant}. The nature of the different regions can be inferred from the analysis of the density of states, magnetization and orbital filling curves.

\begin{figure}
\leavevmode
\includegraphics[clip,width=0.47\textwidth]{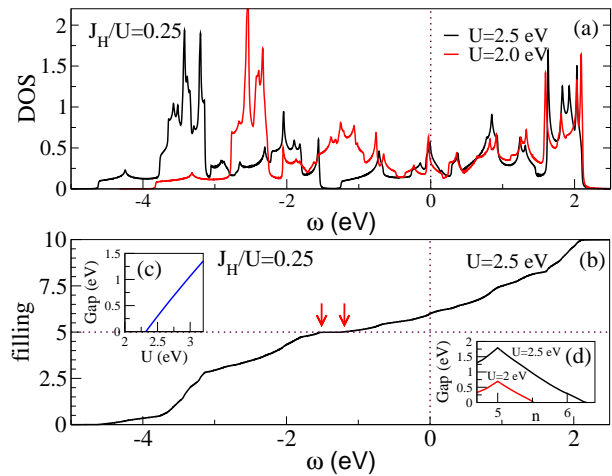}
\caption{(Color online) (a) Total density of states for $J_H/U=0.25$ and 
$U=2$ eV (red) and $U=2.5$ eV (black). The Fermi energy is at $\omega=0$.  
For $U=2.5$ eV, a gap has opened around $\omega \sim -1.5$ eV, which 
corresponds to half-filling as illustrated in (b), where the integrated 
density of states (filling) is shown. This opening of a gap at half-filling 
in the total density of states characterizes the region on the 
right of the dotted curve within the {\it strong orbital differentiation} area 
in Fig.~\ref{fig:pd}. In contrast, there is no gap for $U=2$ eV.  (c) Size 
of the gap at half-filling as a function of $U$ for $n=6$. (d) Gap as a 
function of doping $n$ for the two different values of $U$. The gap has a 
maximum at $n=5$. 
}
\label{fig:dos} 
\end{figure}

We first focus on the  {\it strong orbital differentiation} region of the phase diagram.
Fig.~\ref{fig:dos} represents the total density of states in two
points of this region on both sides of the dotted curve in  Fig.~\ref{fig:pd}: $U=2$~eV and $U=2.5$~eV, both with $J_H/U=0.25$ 
and $n=6$. In both cases, the system is metallic with no gap at the Fermi 
level. However, the two curves are qualitatively different with a gap clearly 
showing below the Fermi energy only for the largest value of $U$.  The opening 
of this gap in the phase diagram is signaled with a  dotted curve in 
Fig.~\ref{fig:pd}. In Fig.~\ref{fig:dos}(b) it is shown that the gap opens at 
an energy that corresponds to half-filling  (five electrons in the five d 
orbitals).   It increases
upon hole doping (decreasing $n$) reaching a maximum at $n=5$
(Fig.~\ref{fig:dos}(d)).
Once this gap opens, its size depends linearly and much stronger on $U$ than the splittings at other energies (Fig.~\ref{fig:dos}(c)). This gap was found and discussed on the basis of LDA+U calculations.\cite{ku2012} 

\begin{figure}
\leavevmode
\includegraphics[clip,width=0.47\textwidth]{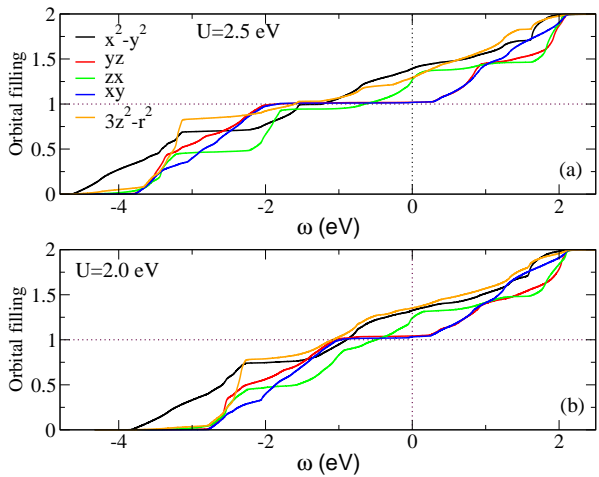}
\caption{(Color online) Orbital fillings as a function of energy  
in the {\it strong orbital differentiation} regime
for (a) $U=2.5$ eV and (b) $U=2$ eV. For $U=2.5$ eV all the orbitals 
show the gap at negative energies. 
$yz$ and $xy$ are gapped at the Fermi energy  while the other three orbitals 
are gapped only below the Fermi energy. 
In contrast, for $U=2$ eV only two orbitals ($yz$ and $xy$) show the gap 
at half-filling while the other three orbitals 
are itinerant.  
}
\label{fig:dos-per-orbital} 
\end{figure}

More can be learned about this {\it strong orbital differentiation} region by looking at the projection of the 
density of states on the five Fe d orbitals. The integrated density of states (filling) of the orbitals as a function of the energy 
is shown in Fig.~\ref{fig:dos-per-orbital} for the same 
two points of the phase diagram as in Fig.~\ref{fig:dos}. 
The first thing we notice is that for both values of the interaction
two orbitals ($xy$ and $yz$) open a gap at half-filling: their gap is already quite large for $U=2$ eV and extends up to the Fermi 
level. 
This orbital selective gap is not apparent when looking at  
the total density of states, Fig.~\ref{fig:dos}. The opening of the gap 
at negative energies in the total density of states is then 
related to the appearance of the gap 
in the other three orbitals $zx$, $x^2-y^2$, and $3z^2-r^2$. 
However, these three orbitals, unlike $xy$ and $yz$, remain itinerant in this region, with a 
finite density of states at the Fermi energy. The orbital differentiation just discussed is concomitant with
orbital ordering: $zx$ tends to be 
more filled and away from half-filling than $yz$, which is stuck to 
half-filling.

\begin{figure}
\leavevmode
\includegraphics[clip,width=0.47\textwidth]{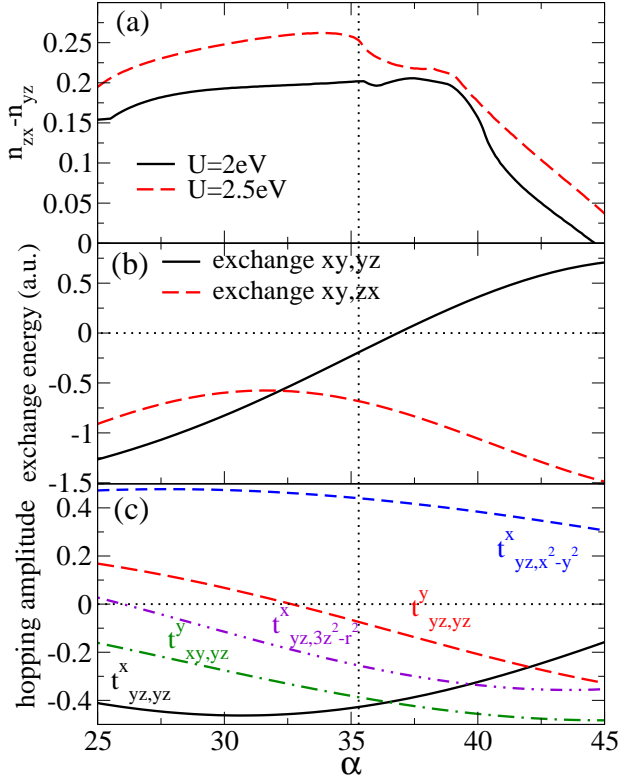}
\caption{
(Color online) Orbital ordering $n_{zx}-n_{yz}$, exchange energy, and the relevant hopping amplitudes as a function of $\alpha$, the angle formed by the Fe-As 
bond and the Fe plane. $\alpha=35.3^o$ (highlighted by a vertical dotted line) 
corresponds to regular, $\alpha> 35.3^o$ to elongated, and 
$\alpha < 35.3^o$ to squeezed tetrahedra. (a) Orbital ordering for different 
values of $U$. It gets supressed with increasing $\alpha$ but it never changes 
sign within the experimentally relevant values of $\alpha$. (b) Exchange 
energy in the $(\pi,0)$ state  
calculated assuming localization of $xy$ and $yz$ (black curve) and 
localization of $xy$ and $zx$ (red curve). Both intra and interorbital exchange contributions are included. As the tetrahedra elongates, the exchange energy would favor a localization of $zx$ versus 
localization of $yz$, but this does not happen, see text for discussion. (c) 
Relevant hopping amplitudes. Note that, by symmetry 
$t^y_{zx,zx}=t^x_{yz,yz}$, $t^y_{zx,x^2-y^2}=t^x_{yz,x^2-y^2}$, 
$t^y_{zx,3z^2-r^2}=t^x_{yz,3z^2-r^2}$, etc.\cite{nosotrasprb09}
}
\label{fig:SE-vs-itinerant} 
\end{figure}

Motivated by this orbital differentiation we analyze the stability of magnetism within a model of localized and itinerant orbitals.  Assuming localized xy and itinerant $x^2-y^2$ and $3z^2-r^2$, we explore the different tendencies of $yz$ and $zx$. To second order in perturbation theory in the hoppings,  
$yz$ ($zx$) has a larger intraorbital exchange along the 
antiferromagnetic $x$-direction (ferromagnetic $y$-direction) favoring localization of $yz$ (delocalization of $zx$) in a 
magnetic state with $(\pi,0)$ ordering.\cite{leeyinku09}
This anisotropic
exchange comes from the counter-intuitive hoppings relation 
$|t^x_{yz,yz}| > |t^y_{yz,yz}|$ arising from the combination of direct Fe-Fe 
and indirect Fe-Pn-Fe hopping amplitudes.\cite{nosotrasprb09}
On the other hand, exchange between $xy$ and $yz$ ($zx$) is 
finite only in the ferromagnetic $y$-direction (antiferromagnetic $x$-direction) and opposes the localization of $yz$. As a consequence, as shown
in Fig.~\ref{fig:SE-vs-itinerant}(b),  there is more energy gain due to the localization of $zx$ than to the localization of $yz$ for regular tetrahedra ($\alpha=35.3^o$). For elongated tetrahedra ($\alpha > 35.3^o$), the localization of $zx$ becomes 
much more advantageous in terms of exchange energy, but this does not affect the sign of the orbital ordering calculated within Hartree-Fock and shown in Fig.~\ref{fig:SE-vs-itinerant}(a). Moreover, the trend changes for smaller values of $\alpha$, where the localization of $yz$ brings an energy gain, but this is not reflected in the magnitude of the orbital ordering in Fig.~\ref{fig:SE-vs-itinerant}(a).

A deeper analysis of the orbital dependent hoppings as a function of $\alpha$ helps to clarify the situation.
The hoppings involving $yz$ and $zx$
are anisotropic in the plane,\cite{nosotrasprb09,leeyinku09} see Fig.~\ref{fig:SE-vs-itinerant}(c). 
 Of those, for a regular tetrahedron ($\alpha=35.3^o$), the largest hoppings
in absolute value in the $x$-direction are $t^x_{yz,yz}$, $t^x_{yz,x^2-y^2}$, $t^x_{yz,3z^2-r^2}$, 
and $t^x_{zx,xy}$. By symmetry, simply exchanging $yz \leftrightarrow zx$, 
the largest hoppings in the $y$-direction are  $t^y_{zx,zx}$, 
$t^y_{zx,x^2-y^2}$, $t^y_{zx,3z^2-r^2}$, and $t^y_{yz,xy}$. From these 
relations, we see that three orbitals ($zx$, $x^2-y^2$, and $3z^2-r^2$) prefer to be 
itinerant to gain kinetic energy in the ferromagnetic $y$-direction.  This 
gain in kinetic energy is maintained when the tetrahedra are elongated while 
the localization of $yz$ would become even more unfavourable in terms of exchange 
energy. In squeezed tetrahedra the hopping between $zx$ and $3z^2-r^2$ along the $y$-direction strongly decreases and induces a reduction of the orbital ordering.
Within this model, the orbital ordering, which does not change sign for the experimentally relevant values of $\alpha$, see Fig.~\ref{fig:SE-vs-itinerant}(a), arises due to the interplay of the exchange energy in the $x$-direction due to $yz$ 
localization and the kinetic energy in the $y$-direction due to the itinerancy 
of $zx$, $x^2-y^2$, and $3z^2-r^2$. Depending on $\alpha$, these two factors 
cooperate  (squeezed tetrahedra) or compete (regular or elongated tetrahedra),
with an overall dominance of the kinetic energy gain. Therefore, this kinetic 
energy gain is important to stabilize the $(\pi,0)$ magnetic ordering in the orbital differentiation regime.

\begin{figure}
\leavevmode
\includegraphics[clip,width=0.45\textwidth]{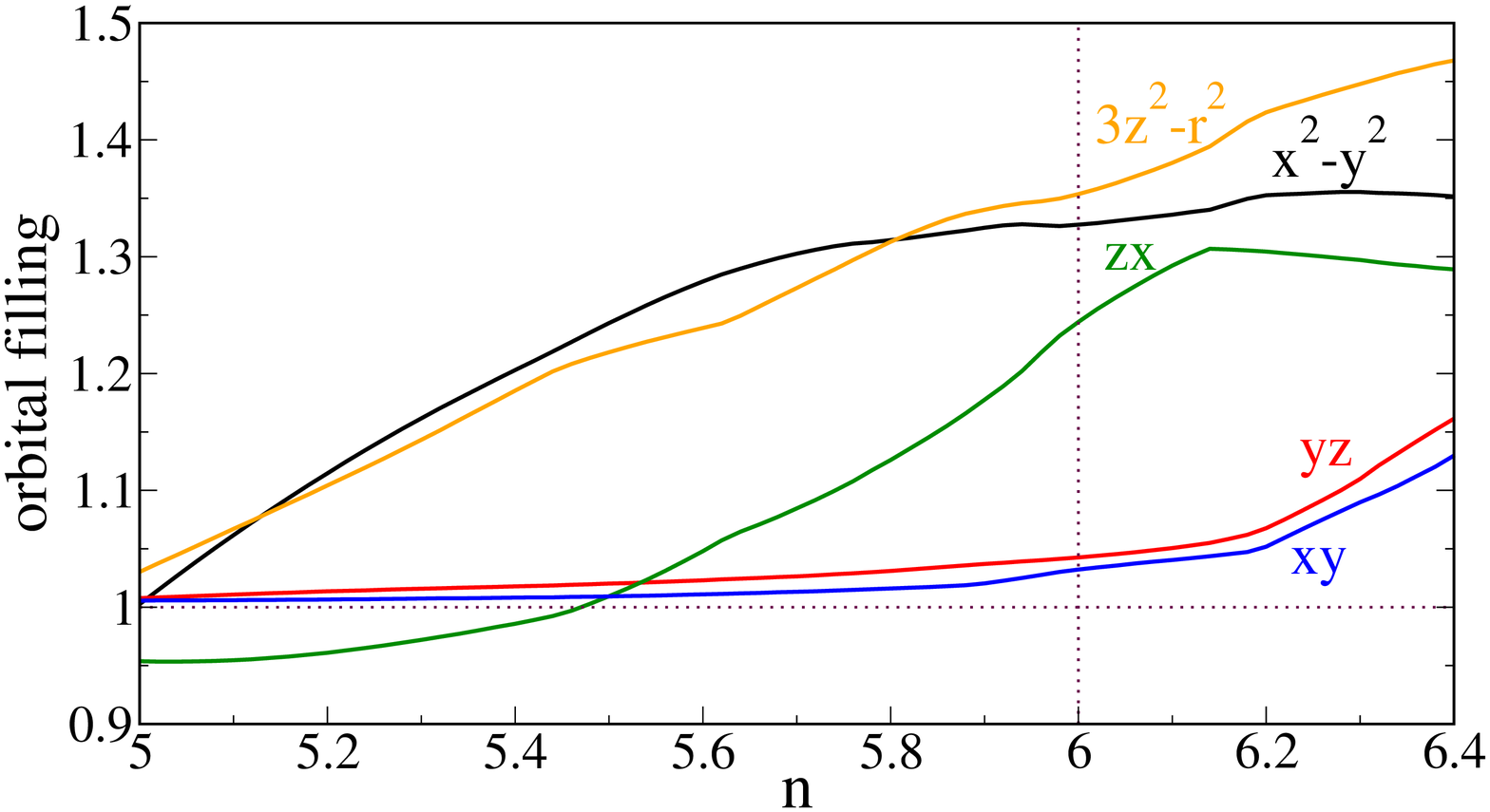}
\caption{
(Color online) Orbital filling as a function of doping $n$ for $U=2$ eV and $J_H/U=0.25$. $yz$ and $xy$ are closer to half filling for all dopings. Doping with holes towards $n=5$ takes all orbitals closer to half-filling but the kinetic energy gain keeps $3z^2-r^2$ and $zx$ itinerant. 
}
\label{fig:filling-vs-n} 
\end{figure}

\begin{figure}
\leavevmode
\includegraphics[clip,width=0.47\textwidth]{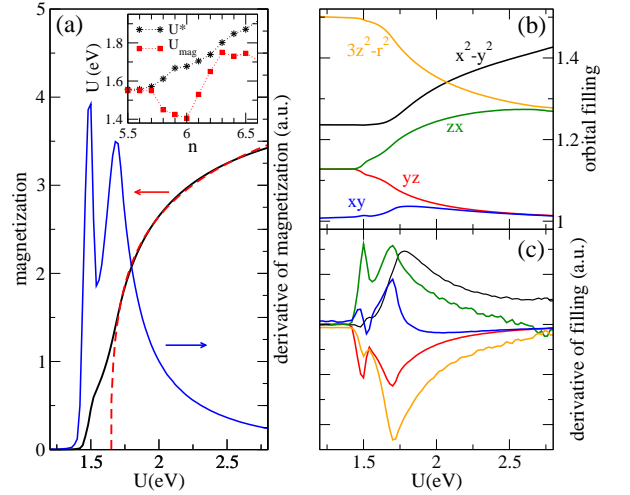}
\caption{
(Color online) (a) Magnetization (black curve) as a function of the Hubbard parameter $U$ for $J_H/U=0.25$. Its derivative (blue, in arbitrary units) has two peaks. The red-dashed curve is a fitting to the concave part of the magnetization which is concomitant with the orbital differentiation. The inset shows the doping dependence of the $U_{mag}$ and $U^*$.  At $U_{mag}$ magnetism appears ($m>0.01$). $U^*$ is the value at which the magnetization changes from a convex to a concave shape, estimated from the point at which the fitted curved intercepts the x-axis.  For values of $U< U^*$, the exact shape of the curve depends on the Fe-As-Fe angle and the tight-binding model, see Fig.~\ref{fig:mag-vs-U}. 
(b) Orbital fillings versus $U$ and (c) their derivatives. At around $U^*$, $xy$, followed by $yz$, tends to half-filling.  
}
\label{fig:mag-filling-derivatives} 
\end{figure}

\begin{figure}
\leavevmode
\includegraphics[clip,width=0.45\textwidth]{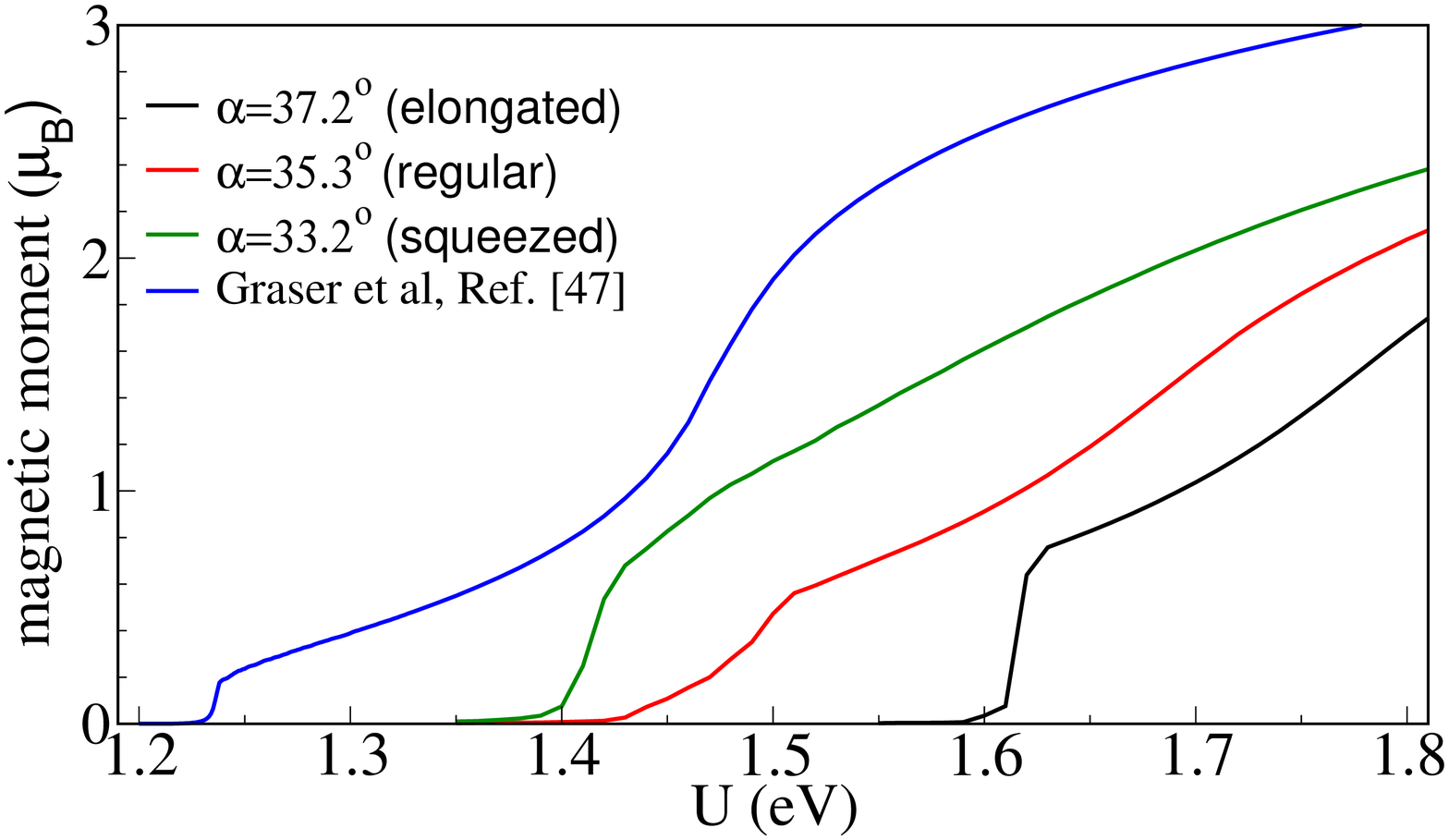}
\caption{
(Color online) Magnetization versus $U$ for $J_H/U=0.25$ for different values 
of the angle $\alpha$ formed by the Fe-As bond with the Fe plane, and for a 
different tight-binding model reported in Ref.~[\onlinecite{graser09}]. For 
sufficiently high values of $U$, the magnetization recovers a typical concave 
shape while for the smaller values of $U$ the curves have different shapes. 
}
\label{fig:mag-vs-U} 
\end{figure}

The orbital differentiation sustained by the gain in kinetic energy in the $y$-direction survives upon doping, as shown in Fig.~\ref{fig:filling-vs-n}. With decreasing $n$ (hole-doping) $zx$ filling decreases fast, changing the sign of the orbital ordering at some point but, remarkably, it is kept away from half-filling, together with $3z^2-r^2$, even at $n=5$.\cite{footnote-n5-pipi}  

 So far, we have discussed the nature of the {\it strong orbital differenciation} region of the phase diagram in Fig.~\ref{fig:pd}.
If we go on decreasing $U$, the orbital differentiation features get weaker, with $xy$ and $yz$ 
getting a finite density 
of states at the Fermi level, and all carriers are itinerant. 

In Fig.~\ref{fig:mag-filling-derivatives} (a) the magnetization is depicted 
as a function of $U$ (black curve)
while the evolution of the orbital fillings and their derivatives with $U$ are 
shown in (b) and (c) respectively.
For the values of $U$ with strong orbital differentiation
the magnetization has the typical concave shape, 
but for smaller values of $U$, the shape of the curve is more complex. 
$xy$ and $yz$ 
orbitals are shown to go to half-filling at around the same value $U^*$ of the 
interactions at which the magnetization becomes concave. Therefore, we estimate the value of $U^*$, which separates the itinerant and strong orbital differentiation regions in Fig. \ref{fig:pd}, from the change of curvature in the 
magnetization, which is concave for $U>U^*$. By fitting this concave part, the 
value of $U^*$ is given by the intercept of the fitting (red-dashed) curve 
with the $x$-axes. $U^*$ is represented by a dashed line in Fig.~\ref{fig:pd}. 

The derivative of the magnetization displayed in 
Fig.~\ref{fig:mag-filling-derivatives}(a) shows two peaks. The second one 
coincides with the appearance of the concave behavior and the tendency of 
$xy$ and $yz$ orbitals to half-filling. The first peak is associated 
with a reorganization of 
the Fermi surface in the ordered state: below,
a spin-density wave between the hole and electron pockets at $(0,0)$ and $(\pi,0)$ respectively characterizes the magnetic state, while above, another spin-density-wave 
instability between the electron pocket at $(0,\pi)$ and the hole pocket 
at $(\pi,\pi)$ gaps this region of the Brillouin zone. This hole pocket is at $(0,0)$ in the 2-Fe Brillouin zone.

Therefore, the detailed behavior of the magnetization in the itinerant 
region with $U< U^*$ depends on the details of the Fermi surface and the tight-binding model and in 
particular on the presence or absence of a hole Fermi pocket at 
$(\pi,\pi)$.  To illustrate this, we show in 
Fig.~\ref{fig:mag-vs-U} the dependence of the magnetization on $U$ for our 
model\cite{nosotrasprb09} with different values of $\alpha$, all with a hole 
Fermi pocket at $(\pi,\pi)$, and for a different tight-binding model\cite{graser09} that 
presents no such Fermi pocket.  The Fermi surfaces for these different 
cases can be found in Refs.~[\onlinecite{nosotrasprb09}] and 
[\onlinecite{graser09}], respectively.  
While both the value of U at which magnetism appears and 
U* depend slightly on the model under consideration, the description presented 
here in terms of itinerant and strong 
orbital differentiation regimes is valid in all these cases. Moreover 
the value of the magnetic moment at the crossover is similar for all these 
models.

The doping dependence of $U^*$ is shown in the inset of Fig.~\ref{fig:mag-filling-derivatives} (a). Its value decreases (increases) upon hole (electron) doping monotonously. This is in contrast with the value of the interaction at which antiferromagnetism appears $U_{mag}$. The different dependence on doping of $U_{mag}$ and $U^*$ gives support to the existence of two regimes: itinerant and strong orbital differentiation. Starting with $n=6$ in the itinerant region and doping with holes the system can either enter in the magnetic orbital differentiation region or lose magnetism, depending on the value of $U$. Note that with hole doping $(\pi,\pi)$ correlations start to dominate over $(\pi,0)$,\cite{nosotrasprb2012} possibly affecting the stability of magnetism.

In conclusion, we have found that except for small values of 
the interactions close to the non magnetic boundary, the metallic $(\pi,0)$ 
antiferromagnetic state satisfying Hund's rule is characterized by a strong 
orbital differentiation between half-filled $xy$ and $yz$ orbitals, 
showing a large gap at the Fermi level 
and itinerant $zx$, $3z^2-r^2$ and $x^2-y^2$ orbitals away from half-filling 
and showing a finite density of states at the Fermi level. 

The 
large gap at the Fermi level shown by the half-filled orbitals suggests a connection to
localization. The present approach cannot address such localization but we believe that it will appear in more elaborate calculations.
Within this interpretation, the larger tendency to localization of $xy$ agrees with it being close to
half-filling in the non-interacting bands and thus very sensitive to 
interactions, while $3z^2-r^2$ and $x^2-y^2$ have the largest filling and 
thus more itinerant behavior. This result is consistent with 
DMFT and slave-spin calculations in the paramagnetic 
state.\cite{yin11,liebsch2010,si2012,aichhorn2010,si2012-2,shen-orb-selective2012} The selective localization of $xy$ orbital could have been 
observed already in ARPES measurements on the 122 selenides in the paramagnetic state.\cite{shen-orb-selective2012} Our calculations also
uncover a strong orbital differentiation  
between $yz$ and $zx$
in the antiferromagnetic state where the tetragonal symmetry is broken. We hope our results encourage new polarized ARPES measurements in the $(\pi,0)$ magnetic state.

In a previous work, we found orbital ordering to promote a larger conductivity in the ferromagnetic 
direction except in a striped region of the phase diagram close to the 
magnetic transition (see Fig. 1(b) in Ref.[\onlinecite{nosotrasprl10_2}]). 
This is consistent with the stabilization of the  ($\pi,0$) magnetic state 
driven by itinerancy in the $y$-direction (and the concomitant orbital order) 
in the region of the phase diagram with both itinerant and 
gapped carriers. 

Current estimates for the interactions\cite{miyake2010} situate the iron 
superconductors close to the boundary between the {\it itinerant} and the 
{\it strong orbital differentiation} regimes. This is in accord with the values of the 
magnetic moment that we obtain in this region, expected in our approximation 
to be similar to those found in LDA calculations.\cite{mazin08-2} 
Moreover, the anisotropy of the conductivity found experimentally in the 122 
compounds,\cite{chu10-1,mazin10} which is such that the conductivity is larger in the antiferromagnetic 
direction, is consistent with the system not being very deep in the orbital 
differentiated regime, where the opposite sign of the anisotropy has been 
calculated.\cite{nosotrasprl10_2}
The exact position in the phase diagram could be different among families.

Finally, the boundary between the {\it itinerant} and the 
{\it strong orbital differentiation} regimes shifts to lower (larger) values of the 
interaction  with hole (electron) doping. Electron doping promotes itinerancy while hole-doping can induce selective  orbital localization. Therefore, by changing the doping 
it could be possible to cross this boundary and enter a different regime. This is consistent with recent DMFT 
calculations\cite{demedici2012} in FeSe
in the non-magnetic state which show a 
selective Mott transition of the $xy$ orbital with hole doping. Slave-spin calculations in 122 
selenides also show that the critical U for the $xy$ selective Mott transition increases with electron doping.\cite{shen-orb-selective2012,si2012-2} In KFe$_2$As$_2$, the $xy$ orbital shows a mass enhancement $\sim 10$, much larger than in undoped 122 pnictides.\cite{shen2012} A change in the sign of the resistivity anisotropy in the magnetic state with hole doping has been recently reported.\cite{sign-reversal2012} Whether this is caused by 
hole-doping induced orbital differentiation is at present not known.

We have benefited from conversations with A. Millis, L. Boeri, I. Eremin, 
M. Capone, L. de Medici, and W. Ku. We acknowledge funding from MINECO-Spain through Grants FIS2008-00124, FIS2009-08744 and FIS2011-29689.

\bibliography{pnictides}

\begin{thebibliography}{51}
\bibitem[{foo()}]{footnote-n5-pipi}
\bibinfo{note}{Note that this is a consequence of the imposed $(\pi,0)$
  ordering. In fact, a $(\pi,\pi)$ checkerboard magnetic order is stabilized
  when heavily doping with holes towards $n=5$. See
  Ref.[\onlinecite{nosotrasprb2012}].}
\end{thebibliography}

\end{document}